\documentclass[aps,prl,twocolumn,superscriptaddress,floatfix,notitlepage]{revtex4-1}

\usepackage{amsthm}

\usepackage{graphicx,psfrag,color}% Include figure files
\usepackage{mathbbol,amsmath,amsfonts,amssymb,bm,dsfont,braket}
\usepackage{wasysym}
\usepackage[section]{placeins}
\usepackage{multirow}

\usepackage{verbatim}

\usepackage{array}
\newcolumntype{x}[1]{>{\centering\hspace{0pt}}p{#1}}
\usepackage[normalem]{ulem}

\begin{document}

\title{Effective Theory of Fermion Quartet Condensation}

\begin{abstract}
We develop a theory of superconductivity (or superfluidity) based on condensed fermion quartets focusing on the dilute spin-$\frac{1}{2}$ systems at zero temperature. 
In the spirit of the Bardeen--Cooper--Schrieffer ansatz, a variational wavefunction is constructed such that, within the so-called ``dilute quartet approximation", it is the ground state of an effective quartic Hamiltonian. 
For a given two-body interaction in favor of quartetting, the gap parameter is suitably defined and the gap equation is also derived. 
As to the excited states, an intuitive physical picture based on a sixteen-dimensional ``occupation space" is depicted and the associated eigen-energies are obtained. 
This theory is applied to compute the superfluid fraction, which is found to be the same as in conventional superconductors, despite the interacting nature of the quartet problem.
\end{abstract}

\author{Qiao-Ru Xu}
\affiliation{Institute for Theoretical Sciences, Westlake University, Hangzhou 310030, China}
\affiliation{Institute of Natural Sciences, Westlake Institute for Advanced Study, Hangzhou 310024, China}

\author{Congjun Wu}
\affiliation{New Cornerstone Science Laboratory, Department of Physics, School of Science, Westlake University, Hangzhou 310030, China}
\affiliation{Institute for Theoretical Sciences, Westlake University, Hangzhou 310030, China}
\affiliation{Institute of Natural Sciences, Westlake Institute for Advanced Study, Hangzhou 310024, China}
\affiliation{Key Laboratory for Quantum Materials of Zhejiang Province, School of Science, Westlake University, Hangzhou 310030, China}

\date{\today}
\maketitle

{\it Introduction.---}The Bardeen--Cooper--Schrieffer (BCS) theory \cite{BCS,Schrieffer} has proved a great success in explaining the conventional superconductivity. 
The building block of the theory is the Cooper pairing \cite{Cooper} of two electrons forming a spin-singlet quasi-bound state with total momentum zero. 
Condensed Cooper pairs constitute the superconducting ground state, described by the BCS variational wavefunction, which is exact when working with an effective quadratic (mean-field) Hamiltonian. 
As a trade-off, the particle number is no longer conserved; rather, the superconducting order parameter, characterized by an energy gap, breaks the $U(1)$ symmetry to $Z^{\,}_2$. 
Since the formulation of the BCS theory, numerous efforts have been made to go beyond the formalism. For example, it was soon generalized to systems with overlapping bands \cite{Moskalenko,Suhl}, where there exists at least two gap parameters, and has been in the spotlight after the discovery of superconductivity at 39 K in MgB$^{\,}_2$ \cite{MgB2}. Another attempt is to develop a theory based on spin-triplet Cooper pairs \cite{AM,BW,AB}, which was found to describe the superfluidity in liquid $^3$He below 3 mK \cite{Leggett1975}. Theories concerning Cooper pairs with finite total momenta are also possible \cite{FF,LO} and might be tested in heavy fermion and cold atom systems \cite{FFLO1,FFLO2}.

Instead of generalized Cooper pairs, multi-particle bound states could also serve as ingredients for unconventional superconductivity or superfluidity. As one of the most promising candidates, condensed fermion quartets (sextets as well) have recently attracted considerable attention \cite{Zhou,Zhang,Han,Pan,Yu,Varma,Lin} owing to the observation of magnetoresistance oscillations with periods being fractions of the flux quantum $\frac{h}{2e}$ in the kagome metal CsV$^{\,}_3$Sb$^{\,}_5$ \cite{Ge}. Historically, among various others \cite{Kivelson1990,Nozieres,Moon,Kivelson2020,Will,Cui}, two mechanisms for the quartet condensation were widely explored, one as quartetting through multicomponent fermions like spin-$\frac{3}{2}$ systems \cite{Schlottmann1994,Wu2005,Lecheminant2005,Lecheminant2009,Guan2009,Schlottmann2012,Solyom2017,Neupert2024} and the other as an intertwined vestigial order of a primary phase \cite{Volovik1989,Vishwanath2009,Kivelson2009,Sudbo2010,Fernandes2021,Yao2021,Wu2024,Wu2023,Fernandes2023,Wang2024}. Despite a wealth of progress on quartetting mechanisms, only solvable models were designed \cite{Wang2022,Hu2024} to uncover physical properties of condensed quartets while a general theory is still missing and highly desirable. In this paper, focusing mainly on dilute spin-$\frac{1}{2}$ systems at zero temperature, we develop an effective theory of condensed fermion quartets, in the spirit of BCS, and apply it to calculate the superfluid fraction. Due to the intrinsic difficulty of solving a nontrivial interacting problem, we start by discussing an approximation that we adopted.

\begin{figure}[b]
	\includegraphics[width=\columnwidth]{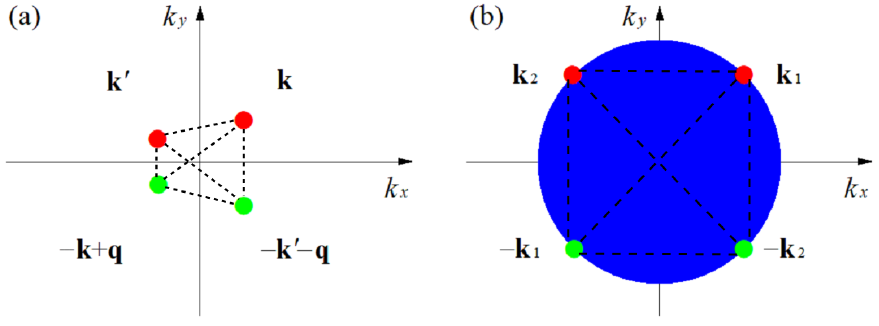}
	\caption{Schematic plots of two different scenarios for the fermion quartet formation in spin-$\frac{1}{2}$ systems, with red solid circles representing spin $\uparrow$ and green ones spin $\downarrow$. (a) Binding of two tightly bound Cooper pairs with opposite total momenta such that the net momentum of the quartet is still zero and (b) binding of two loosely bound Cooper pairs (each with total momentum zero) on the Fermi surface (in blue).}\label{Fig_Binding}
\end{figure}

{\it Wavefunction and dilute quartet approximation.---}A zero momentum Cooper pair is a superposition of the product of two electrons with opposite momenta $\pm\mathbf{k}$'s. A natural extension to a zero momentum fermion quartet is a superposition of the product of two Cooper pairs with opposite momenta $\pm\mathbf{q}$'s [see Fig.\,\ref{Fig_Binding} (a)]. Although focusing mainly on the dilute regime, at the end of the paper we will briefly discuss the scenario in the presence of a Fermi surface with $\mathbf{q}=0$ fixed [see Fig.\,\ref{Fig_Binding} (b)]. Now, in the spirit of the BCS ansatz, we construct a general variational wavefunction of condensed fermion quartets as follows
\begin{align}\label{w.f.}
    \ket{\Psi}=\frac{1}{\sqrt{\mathcal{N}}} \prod^{\,}_{(\mathbf{k}\mathbf{k'}\mathbf{q})}(1+\alpha^{\,}_{\mathbf{k}\mathbf{k'}\mathbf{q}}Q^\dagger_{\mathbf{k}\mathbf{k'}\mathbf{q}})\ket{0},
\end{align}
where $\mathcal{N}$ is the normalization factor, $\alpha^{\,}_{\mathbf{k}\mathbf{k'}\mathbf{q}}$ is the variational parameter, $Q^\dagger_{\mathbf{k}\mathbf{k'}\mathbf{q}}=c^\dagger_{\mathbf{k}\uparrow}c^\dagger_{-\mathbf{k}+\mathbf{q}\downarrow}c^\dagger_{\mathbf{k'}\uparrow}c^\dagger_{-\mathbf{k'}-\mathbf{q}\downarrow}$, with $c^\dagger_{\mathbf{k}\sigma}$ the creation operator of spin-$\frac{1}{2}$ fermions of momentum $\mathbf{k}$ and spin $\sigma \in \{\uparrow,\downarrow\}$, and $\ket{0}$ is the vacuum of the annihilation operator $c^{\,}_{\mathbf{k}\sigma}$. Because $\alpha^{\,}_{\mathbf{k}\mathbf{k'}\mathbf{q}}$ satisfies the symmetry relations
\begin{align}\label{symm}
    \alpha^{\,}_{\mathbf{k}\mathbf{k'}\mathbf{q}}=\alpha^{\,}_{\mathbf{k'},\mathbf{k},\mathbf{-q}}=-\alpha^{\,}_{\mathbf{k},\mathbf{k'},\mathbf{k-k'-q}}=-\alpha^{\,}_{\mathbf{k'},\mathbf{k},\mathbf{k'-k+q}},    
\end{align}
the four brackets $(\mathbf{k}\mathbf{k'}\mathbf{q})$, $(\mathbf{k'},\mathbf{k},\mathbf{-q})$, $(\mathbf{k},\mathbf{k'},\mathbf{k-k'-q})$, and $(\mathbf{k'},\mathbf{k},\mathbf{k'-k+q})$ will be counted once in the product $\prod^{\,}_{(\mathbf{k}\mathbf{k'}\mathbf{q})}$ to avoid double counting. Therefore, any two brackets from the product $\prod^{\,}_{(\mathbf{k}\mathbf{k'}\mathbf{q})}$ should be identified as different.

For dilute many-body systems with the momentum occupation number $n^{\,}_{\mathbf{k}\sigma}\ll 1$, if we choose two brackets $(\mathbf{k}^{\,}_1\mathbf{k}^{\,}_2\mathbf{q}^{\,}_1)$ and $(\mathbf{k}^{\,}_3\mathbf{k}^{\,}_4\mathbf{q}^{\,}_2)$ from the product $\prod^{\,}_{(\mathbf{k}\mathbf{k'}\mathbf{q})}$, it is appropriate to make an approximation when computing the commutator
\begin{align}
	[Q^{\,}_{\mathbf{k}^{\,}_1\mathbf{k}^{\,}_2\mathbf{q}^{\,}_1},\, Q^\dagger_{\mathbf{k}^{\,}_3\mathbf{k}^{\,}_4\mathbf{q}^{\,}_2}]\simeq 0,
\end{align}
which we will call the ``dilute quartet approximation" (DQA) and will be repeatedly used later. To see the validity of this approximation, let us look at a particular example with $\mathbf{k}^{\,}_3=\mathbf{k}^{\,}_1$ and $\mathbf{k}^{\,}_4=\mathbf{k}^{\,}_2$ but $\mathbf{q}^{\,}_2\neq \mathbf{q}^{\,}_1$. Note that by construction we have $\mathbf{k}^{\,}_1\neq \mathbf{k}^{\,}_2$ and $\mathbf{q}^{\,}_2\neq \mathbf{k}^{\,}_1-\mathbf{k}^{\,}_2-\mathbf{q}^{\,}_1$. Then it is straightforward to find that
\begin{align}\label{commutator}
    [Q^{\,}_{\mathbf{k}^{\,}_1\mathbf{k}^{\,}_2\mathbf{q}^{\,}_1},\, Q^\dagger_{\mathbf{k}^{\,}_1\mathbf{k}^{\,}_2\mathbf{q}^{\,}_2}]=&- P^{\,}_{\mathbf{k}^{\,}_1\mathbf{k}^{\,}_2\mathbf{q}^{\,}_1}P^\dagger_{\mathbf{k}^{\,}_1\mathbf{k}^{\,}_2\mathbf{q}^{\,}_2}\nonumber\\
    &+T^{\,}_{\mathbf{k}^{\,}_1\mathbf{k}^{\,}_2\mathbf{q}^{\,}_1}T^\dagger_{\mathbf{k}^{\,}_1\mathbf{k}^{\,}_2\mathbf{q}^{\,}_2}\nonumber\\
    &+ T^{\,}_{\mathbf{k}^{\,}_2,\mathbf{k}^{\,}_1,-\mathbf{q}^{\,}_1}T^\dagger_{\mathbf{k}^{\,}_2,\mathbf{k}^{\,}_1,-\mathbf{q}^{\,}_2},
\end{align}
where we have defined $P^\dagger_{\mathbf{k}\mathbf{k'}\mathbf{q}}=c^\dagger_{-\mathbf{k}+\mathbf{q}\downarrow}c^\dagger_{-\mathbf{k'}-\mathbf{q}\downarrow}$ and $T^\dagger_{\mathbf{k}\mathbf{k'}\mathbf{q}}=c^\dagger_{\mathbf{k}\uparrow}c^\dagger_{-\mathbf{k}+\mathbf{q}\downarrow}c^\dagger_{-\mathbf{k'}-\mathbf{q}\downarrow}$. As we will see later, $P^\dagger_{\mathbf{k}^{\,}_1\mathbf{k}^{\,}_2\mathbf{q}^{\,}_1}\ket{\Psi}$ and $P^\dagger_{\mathbf{k}^{\,}_1\mathbf{k}^{\,}_2\mathbf{q}^{\,}_2}\ket{\Psi}$ are different two-particle excited states while $T^\dagger_{\mathbf{k}^{\,}_1\mathbf{k}^{\,}_2\mathbf{q}^{\,}_1}\ket{\Psi}$, $T^\dagger_{\mathbf{k}^{\,}_1\mathbf{k}^{\,}_2\mathbf{q}^{\,}_2}\ket{\Psi}$, $T^\dagger_{\mathbf{k}^{\,}_2,\mathbf{k}^{\,}_1,-\mathbf{q}^{\,}_1}\ket{\Psi}$, and $T^\dagger_{\mathbf{k}^{\,}_2,\mathbf{k}^{\,}_1,-\mathbf{q}^{\,}_2}\ket{\Psi}$ are different three-particle excited states. Because different excited states are orthogonal to each other, the expectation value of the commutator in Eq.\,(\ref{commutator}) vanishes. This explains the physics behind the DQA. Now we can compute the normalization factor as $\mathcal{N}= \prod^{\,}_{(\mathbf{k}\mathbf{k'}\mathbf{q})}(1+|\alpha^{\,}_{\mathbf{k}\mathbf{k'}\mathbf{q}}|^2)$ after employing the approximation.

{\it Effective Hamiltonian and gap equation.---}Within the DQA, $\ket{\Psi}$ is the ground state of the effective quartic Hamiltonian
\begin{align}
\widehat{\mathcal{H}}=&\sum_{\mathbf{k}\sigma}\xi^{\,}_{\mathbf{k}}c^\dagger_{\mathbf{k}\sigma} c^{\,}_{\mathbf{k}\sigma}+\sum^{\,}_{(\mathbf{k}\mathbf{k'}\mathbf{q})}(\Delta^{\,}_{\mathbf{k}\mathbf{k'}\mathbf{q}}Q^\dagger_{\mathbf{k}\mathbf{k'}\mathbf{q}}+\text{H.c.})\nonumber\\
&-\mathfrak{Re}\sum^{\,}_{(\mathbf{k}\mathbf{k'}\mathbf{q})}\Delta^{\,}_{\mathbf{k}\mathbf{k'}\mathbf{q}}\braket{Q^\dagger_{\mathbf{k}\mathbf{k'}\mathbf{q}}},
\end{align}
where $\xi^{\,}_{\mathbf{k}}=\epsilon^{\,}_{\mathbf{k}}-\mu$ is the kinetic energy $\epsilon^{\,}_{\mathbf{k}}=\frac{\hbar^2k^2}{2m}$ (with $k=|\mathbf{k}|$) measured from the chemical potential $\mu$ and $\Delta^{\,}_{\mathbf{k}\mathbf{k'}\mathbf{q}}$ is the quartetting gap parameter with the symmetry relations 
\begin{align}\label{symmetry}
    \Delta^{\,}_{\mathbf{k}\mathbf{k'}\mathbf{q}}=\Delta^{\,}_{\mathbf{k'},\mathbf{k},\mathbf{-q}}=-\Delta^{\,}_{\mathbf{k},\mathbf{k'},\mathbf{k-k'-q}}=-\Delta^{\,}_{\mathbf{k'},\mathbf{k},\mathbf{k'-k+q}},
\end{align}
same as Eq.\,(\ref{symm}) for $\alpha^{\,}_{\mathbf{k}\mathbf{k'}\mathbf{q}}$. For a given two-body interaction $\sum^{\,}_{\mathbf{k}\mathbf{k'}\mathbf{q}}V^{\,}_{\mathbf{k}\mathbf{k'}}c^\dagger_{\mathbf{k}\uparrow}c^\dagger_{-\mathbf{k}+\mathbf{q}\downarrow}c^{\,}_{-\mathbf{k'}+\mathbf{q}\downarrow}c^{\,}_{\mathbf{k'}\uparrow}$ that favors the fermion quartet formation, we can define the gap parameter as follows 
\begin{align}\label{def2}
    \Delta^{\,}_{\mathbf{k}\mathbf{k'}\mathbf{q}}=\sum_\mathbf{k''}\bigg[&(V^{\,}_{\mathbf{k}\mathbf{k''}}\braket{Q^{\,}_{\mathbf{k''}\mathbf{k'}\mathbf{q}}}+V^{\,}_{\mathbf{k'}\mathbf{k''}}\braket{Q^{\,}_{\mathbf{k''},\mathbf{k},-\mathbf{q}}})\nonumber\\
    &-(\mathbf{q}\rightarrow\mathbf{k}-\mathbf{k'}-\mathbf{q})\bigg],
\end{align}
which satisfies the symmetry relations Eq.\,(\ref{symmetry}). After some calculation, we obtain the total energy
\begin{align}\label{totalenergy}
\braket{\widehat{\mathcal{H}}}=\sum^{\,}_{(\mathbf{k}\mathbf{k'}\mathbf{q})}\frac{2\xi^{\,}_{\mathbf{k}\mathbf{k'}\mathbf{q}}|\alpha^{\,}_{\mathbf{k}\mathbf{k'}\mathbf{q}}|^2+\mathfrak{Re}(\Delta^{\,}_{\mathbf{k}\mathbf{k'}\mathbf{q}}\alpha^*_{\mathbf{k}\mathbf{k'}\mathbf{q}})}{1+|\alpha^{\,}_{\mathbf{k}\mathbf{k'}\mathbf{q}}|^2},
\end{align}
with $2\xi^{\,}_{\mathbf{k}\mathbf{k'}\mathbf{q}}=\xi^{\,}_{\mathbf{k}}+\xi^{\,}_{-\mathbf{k}+\mathbf{q}}+\xi^{\,}_{\mathbf{k'}}+\xi^{\,}_{-\mathbf{k'}-\mathbf{q}}$. 
Minimizing the total energy, we get the variational parameter 
\begin{align}\label{varationalparameter}
\alpha^{\,}_{\mathbf{k}\mathbf{k'}\mathbf{q}}=\frac{\xi^{\,}_{\mathbf{k}\mathbf{k'}\mathbf{q}}-E^{\,}_{\mathbf{k}\mathbf{k'}\mathbf{q}}}{\Delta^*_{\mathbf{k}\mathbf{k'}\mathbf{q}}},
\end{align}
with $E^{\,}_{\mathbf{k}\mathbf{k'}\mathbf{q}}=\sqrt{\xi^2_{\mathbf{k}\mathbf{k'}\mathbf{q}}+|\Delta^{\,}_{\mathbf{k}\mathbf{k'}\mathbf{q}}|^2}$. %Combing Eqs.\,(\ref{gapparameter}) and (\ref{varationalparameter}), 
Finally, we arrive at the gap equation
\begin{align}\label{gapequation}
\Delta^{\,}_{\mathbf{k}\mathbf{k'}\mathbf{q}}=-\sum^{\,}_\mathbf{k''}\bigg[&(\frac{V^{\,}_{\mathbf{k}\mathbf{k''}}\Delta^{\,}_{\mathbf{k''}\mathbf{k'}\mathbf{q}}}{2E^{\,}_{\mathbf{k''}\mathbf{k'}\mathbf{q}}}+\frac{V^{\,}_{\mathbf{k'}\mathbf{k''}}\Delta^{\,}_{\mathbf{k''},\mathbf{k},-\mathbf{q}}}{2E^{\,}_{\mathbf{k''},\mathbf{k},-\mathbf{q}}})\nonumber\\
&-(\mathbf{q}\rightarrow\mathbf{k}-\mathbf{k'}-\mathbf{q})\bigg].
\end{align}

By the way, effective few-body interactions such as three- and four-body interactions are often featured in atomic and nuclear physics \cite{Hammer}. Then from the mean-field perspective we can define the gap parameter as 
\begin{align}\label{def4}
    \Delta^{\,}_{\mathbf{k}^{\,}_1\mathbf{k}^{\,}_2\mathbf{q}^{\,}_1}=\sum^{\,}_{(\mathbf{k}^{\,}_3\mathbf{k}^{\,}_4\mathbf{q}^{\,}_2)} V^{\,}_{\mathbf{k}^{\,}_1\mathbf{k}^{\,}_2\mathbf{q}^{\,}_1\mathbf{k}^{\,}_3\mathbf{k}^{\,}_4\mathbf{q}^{\,}_2}\braket{Q^{\,}_{\mathbf{k}^{\,}_3\mathbf{k}^{\,}_4\mathbf{q}^{\,}_2}},
\end{align}
where the effective four-body (quartetting) interaction $V^{\,}_{\mathbf{k}^{\,}_1\mathbf{k}^{\,}_2\mathbf{q}^{\,}_1\mathbf{k}^{\,}_3\mathbf{k}^{\,}_4\mathbf{q}^{\,}_2}$ is introduced in the following form $\sum^{\,}_{(\mathbf{k}^{\,}_1\mathbf{k}^{\,}_2\mathbf{q}^{\,}_1)}\sum^{\,}_{(\mathbf{k}^{\,}_3\mathbf{k}^{\,}_4\mathbf{q}^{\,}_2)} V^{\,}_{\mathbf{k}^{\,}_1\mathbf{k}^{\,}_2\mathbf{q}^{\,}_1\mathbf{k}^{\,}_3\mathbf{k}^{\,}_4\mathbf{q}^{\,}_2}Q^\dagger_{\mathbf{k}^{\,}_1\mathbf{k}^{\,}_2\mathbf{q}^{\,}_1}Q^{\,}_{\mathbf{k}^{\,}_3\mathbf{k}^{\,}_4\mathbf{q}^{\,}_2}$. If the quartetting interaction $V^{\,}_{\mathbf{k}^{\,}_1\mathbf{k}^{\,}_2\mathbf{q}^{\,}_1\mathbf{k}^{\,}_3\mathbf{k}^{\,}_4\mathbf{q}^{\,}_2}=V$ has no momentum dependence, the same will be true for the gap parameter $\Delta^{\,}_{\mathbf{k}^{\,}_1\mathbf{k}^{\,}_2\mathbf{q}^{\,}_1}=\Delta$, which differs from the gap parameter defined through the two-body interaction $V^{\,}_{\mathbf{k}\mathbf{k'}}$ in Eq.\,(\ref{def2}). Finally, we get the same Eqs.\,(\ref{totalenergy}) and (\ref{varationalparameter}), but with Eq.\, (\ref{gapequation}) modified
\begin{align}
\Delta^{\,}_{\mathbf{k}^{\,}_1\mathbf{k}^{\,}_2\mathbf{q}^{\,}_1}=-\sum^{\,}_{(\mathbf{k}^{\,}_3\mathbf{k}^{\,}_4\mathbf{q}^{\,}_2)}\frac{V^{\,}_{\mathbf{k}^{\,}_1\mathbf{k}^{\,}_2\mathbf{q}^{\,}_1\mathbf{k}^{\,}_3\mathbf{k}^{\,}_4\mathbf{q}^{\,}_2}\Delta^{\,}_{\mathbf{k}^{\,}_3\mathbf{k}^{\,}_4\mathbf{q}^{\,}_2}}{2E^{\,}_{\mathbf{k}^{\,}_3\mathbf{k}^{\,}_4\mathbf{q}^{\,}_2}}.
\end{align}

{\it Single- and multi-particle excitations.---}With the variational parameter given by Eq.\,(\ref{varationalparameter}), a direct calculation within the DQA (i.e., $\widehat{\mathcal{H}}\ket{\Psi}=E^{\,}_0\ket{\Psi}$) reveals the ground state eigen-energy
\begin{align}\label{energy0}
E^{\,}_0=-C+\sum^{\,}_{(\mathbf{k}^{\,}_1\mathbf{k}^{\,}_2\mathbf{q}^{\,}_1)}H^{\,}_{\mathbf{k}^{\,}_1\mathbf{k}^{\,}_2\mathbf{q}^{\,}_1},
\end{align}
where  $C=\mathfrak{Re}\sum^{\,}_{(\mathbf{k}^{\,}_1\mathbf{k}^{\,}_2\mathbf{q}^{\,}_1)}\Delta^{\,}_{\mathbf{k}^{\,}_1\mathbf{k}^{\,}_2\mathbf{q}^{\,}_1}\braket{Q^\dagger_{\mathbf{k}^{\,}_1\mathbf{k}^{\,}_2\mathbf{q}^{\,}_1}}$ is a constant and $H^{\,}_{\mathbf{k}^{\,}_1\mathbf{k}^{\,}_2\mathbf{q}^{\,}_1}=\xi^{\,}_{\mathbf{k}^{\,}_1\mathbf{k}^{\,}_2\mathbf{q}^{\,}_1}-E^{\,}_{\mathbf{k}^{\,}_1\mathbf{k}^{\,}_2\mathbf{q}^{\,}_1}$ is called the hybridization energy that we will see later. By constructing single- and multi-particle excited states, we can obtain the other eigen-energies in a similar way. For example, after constructing a single-particle excited state $c^\dagger_{\mathbf{k}\uparrow}\ket{\Psi}\equiv\ket{\mathbf{k}}$ and a three-particle excited state $c^\dagger_{-\mathbf{k}+\mathbf{q}\downarrow}c^\dagger_{\mathbf{k'}\uparrow}c^\dagger_{-\mathbf{k'}-\mathbf{q}\downarrow}\ket{\Psi}\equiv\ket{\mathbf{k}\mathbf{k'}\mathbf{q}}$, we can do direct calculations $\widehat{\mathcal{H}}\ket{\mathbf{k}}=E^{\,}_1\ket{\mathbf{k}}$ and $\widehat{\mathcal{H}}\ket{\mathbf{k}\mathbf{k'}\mathbf{q}}=E^{\,}_3\ket{\mathbf{k}\mathbf{k'}\mathbf{q}}$ to obtain eigen-energies of the system as follows
\begin{align}
&E^{\,}_1=-C+\xi^{\,}_{\mathbf{k}}+\sum^{\,}_{\substack{(\mathbf{k}^{\,}_1\mathbf{k}^{\,}_2\mathbf{q}^{\,}_1)\\ \mathbf{k}^{\,}_j\neq\mathbf{k}(j=1,2)}}H^{\,}_{\mathbf{k}^{\,}_1\mathbf{k}^{\,}_2\mathbf{q}^{\,}_1},\label{energy1}\\
&E^{\,}_3=-C+(2\xi^{\,}_{\mathbf{k}\mathbf{k'}\mathbf{q}}-\xi^{\,}_{\mathbf{k}})+\hspace*{-1.3cm}\sum^{\,}_{\substack{(\mathbf{k}^{\,}_1\mathbf{k}^{\,}_2\mathbf{q}^{\,}_1)\\ \mathbf{k}^{\,}_j\neq\mathbf{k'}(j=1,2)\\ -\mathbf{k}^{\,}_1+\mathbf{q}^{\,}_1,-\mathbf{k}^{\,}_2-\mathbf{q}^{\,}_1\notin\{-\mathbf{k}+\mathbf{q},-\mathbf{k'}-\mathbf{q}\}}}\hspace*{-1.3cm}H^{\,}_{\mathbf{k}^{\,}_1\mathbf{k}^{\,}_2\mathbf{q}^{\,}_1}.\label{energy3}
\end{align}
By subtracting Eq.\,(\ref{energy0}) from Eqs.\,(\ref{energy1}) and (\ref{energy3}), we arrive at a single-particle excitation energy $\omega^{\,}_\mathbf{k}$ and a three-particle excitation energy $\omega^{\,}_{\mathbf{k}\mathbf{k'}\mathbf{q}}$, respectively.

\begin{table}[t]
	\centering
	\caption{The basis of a sixteen-dimensional ``occupation space" of the four single-particle states $\mathbf{k}\uparrow$, $-\mathbf{k}+\mathbf{q}\downarrow$, $\mathbf{k'}\uparrow$, and $-\mathbf{k'}-\mathbf{q}\downarrow$ associated with the bracket $(\mathbf{k}\mathbf{k'}\mathbf{q})$. The vacuum and the four-particle occupied state are shown in the first row. The second row is associated with the four single-particle occupied states. The six two-particle occupied states and the four three-particle occupied states are presented in the third row and the fourth row, respectively.}
	\smallskip
	\begin{tabular}{c}
		\hline\hline\\[-1ex]
		$\ket{0},\qquad\qquad Q^\dagger_{\mathbf{k}\mathbf{k'}\mathbf{q}}\ket{0}$ \\ [2ex]
		\hline\\ [-1ex]
		$c^\dagger_{\mathbf{k}\uparrow}\ket{0},\qquad\,\,
		c^\dagger_{\mathbf{-k}+\mathbf{q}\downarrow}\ket{0},\qquad\,\,
		c^\dagger_{\mathbf{k'}\uparrow}\ket{0},\qquad\,\,
		c^\dagger_{\mathbf{-k'}-\mathbf{q}\downarrow}\ket{0}$\\ [2ex]
		\hline\\ [-1ex]
		$c^\dagger_{\mathbf{k}\uparrow}c^\dagger_{\mathbf{-k}+\mathbf{q}\downarrow}\ket{0},\qquad\,\,\,
		c^\dagger_{\mathbf{k}\uparrow}c^\dagger_{\mathbf{k'}\uparrow}\ket{0},\qquad\qquad\quad\,\,
		c^\dagger_{\mathbf{k}\uparrow}c^\dagger_{\mathbf{-k'}-\mathbf{q}\downarrow}\ket{0}$,\\ [2ex]
		$c^\dagger_{\mathbf{-k}+\mathbf{q}\downarrow}c^\dagger_{\mathbf{k'}\uparrow}\ket{0},\qquad\,
		c^\dagger_{\mathbf{-k}+\mathbf{q}\downarrow}c^\dagger_{\mathbf{-k'}-\mathbf{q}\downarrow}\ket{0},\qquad\,
		c^\dagger_{\mathbf{k'}\uparrow}c^\dagger_{\mathbf{-k'}-\mathbf{q}\downarrow}\ket{0}$\\ [2ex]
		\hline\\ [-1ex]
		$c^\dagger_{\mathbf{k}\uparrow}c^\dagger_{\mathbf{-k}+\mathbf{q}\downarrow}c^\dagger_{\mathbf{k'}\uparrow}\ket{0},\qquad\quad\,\,
		c^\dagger_{\mathbf{k}\uparrow}c^\dagger_{\mathbf{-k}+\mathbf{q}\downarrow}c^\dagger_{\mathbf{-k'}-\mathbf{q}\downarrow}\ket{0}$,\\ [2ex]
		$c^\dagger_{\mathbf{k}\uparrow}c^\dagger_{\mathbf{k'}\uparrow}c^\dagger_{\mathbf{-k'}-\mathbf{q}\downarrow}\ket{0},\qquad\quad\,
		c^\dagger_{\mathbf{-k}+\mathbf{q}\downarrow}c^\dagger_{\mathbf{k'}\uparrow}c^\dagger_{\mathbf{-k'}-\mathbf{q}\downarrow}\ket{0}$\\ [2ex]
		\hline\hline
	\end{tabular}
	\label{tab:basis}
\end{table}

To build an intuition of these excitations, let us temporarily focus on a particular bracket $(\mathbf{k}\mathbf{k'}\mathbf{q})$ and consider a sixteen-dimensional ``occupation space" \cite{Leggett2022} of the four single-particle states $\mathbf{k}\uparrow$, $-\mathbf{k}+\mathbf{q}\downarrow$, $\mathbf{k'}\uparrow$, and $-\mathbf{k'}-\mathbf{q}\downarrow$ spanned by the basis as shown in Table\,\ref{tab:basis}. In the presence of the quartetting potential $\Delta^{\,}_{\mathbf{k}\mathbf{k'}\mathbf{q}}$, there exists hybridization %(see Fig.\,\ref{Fig_Occupation}) 
between the vacuum $\ket{0}$ and the four-particle occupied state $Q^\dagger_{\mathbf{k}\mathbf{k'}\mathbf{q}}\ket{0}$, with the energy of the hybridized vacuum $\ket{0}+\alpha^{\,}_{\mathbf{k}\mathbf{k'}\mathbf{q}}Q^\dagger_{\mathbf{k}\mathbf{k'}\mathbf{q}}\ket{0}$ being $\xi^{\,}_{\mathbf{k}\mathbf{k'}\mathbf{q}}-E^{\,}_{\mathbf{k}\mathbf{k'}\mathbf{q}}$, which is just $H^{\,}_{\mathbf{k}\mathbf{k'}\mathbf{q}}$ of Eq.\,(\ref{energy0}). Thus, the excitation from the hybridized vacuum to the single-particle occupied state $c^\dagger_{\mathbf{k}\uparrow}\ket{0}$ costs the energy $\xi^{\,}_{\mathbf{k}}-H^{\,}_{\mathbf{k}\mathbf{k'}\mathbf{q}}$. However, to obtain the excitation energy $\omega^{\,}_\mathbf{k}$ associated with the single-particle excited state $\ket{\mathbf{k}}=c^\dagger_{\mathbf{k}\uparrow}\ket{\Psi}$, we need to consider not only one bracket $(\mathbf{k}\mathbf{k'}\mathbf{q})$ but many brackets associated with the single-particle state $\mathbf{k}\uparrow$. Including contributions from all possible brackets, we therefore obtain the single-particle excitation energy $\omega^{\,}_\mathbf{k}$ as follows
\begin{align}
\omega^{\,}_\mathbf{k}=\xi^{\,}_{\mathbf{k}}-
\sum^{\,}_{(\mathbf{k}^{\,}_1\mathbf{k}^{\,}_2\mathbf{q}^{\,}_1)}H^{\,}_{\mathbf{k}^{\,}_1\mathbf{k}^{\,}_2\mathbf{q}^{\,}_1}(\delta^{\,}_{\mathbf{k}^{\,}_1\mathbf{k}}+\delta^{\,}_{\mathbf{k}^{\,}_2\mathbf{k}}),
\end{align}
which is $E^{\,}_1-E^{\,}_0$ as expected from Eqs.\,(\ref{energy0}) and (\ref{energy1}). Similarly, we can obtain the other excitation energies. For example, with the shorthand $\delta^{\,}_{\{\mathbf{k}^{\,}_1,\mathbf{k}^{\,}_2\}\mathbf{k}}\equiv\delta^{\,}_{\mathbf{k}^{\,}_1\mathbf{k}}+\delta^{\,}_{\mathbf{k}^{\,}_2\mathbf{k}}$, the excitation energy associated with the single-particle excited state $c^\dagger_{-\mathbf{k}+\mathbf{q}\downarrow}\ket{\Psi}$ is obtained as $\xi^{\,}_{-\mathbf{k}+\mathbf{q}}-
\sum^{\,}_{(\mathbf{k}^{\,}_1\mathbf{k}^{\,}_2\mathbf{q}^{\,}_1)}H^{\,}_{\mathbf{k}^{\,}_1\mathbf{k}^{\,}_2\mathbf{q}^{\,}_1}\delta^{\,}_{\{-\mathbf{k}^{\,}_1+\mathbf{q}^{\,}_1,-\mathbf{k}^{\,}_2-\mathbf{q}^{\,}_1\}-\mathbf{k}+\mathbf{q}}$, and the excitation energy $\omega^{\,}_{\mathbf{k}\mathbf{k'}\mathbf{q}}$ associated with the three-particle excited state $\ket{\mathbf{k}\mathbf{k'}\mathbf{q}}=c^\dagger_{-\mathbf{k}+\mathbf{q}\downarrow}c^\dagger_{\mathbf{k'}\uparrow}c^\dagger_{-\mathbf{k'}-\mathbf{q}\downarrow}\ket{\Psi}$ is obtained as follows 
\begin{align}
\omega^{\,}_{\mathbf{k}\mathbf{k'}\mathbf{q}}=&(2\xi^{\,}_{\mathbf{k}\mathbf{k'}\mathbf{q}}-\xi^{\,}_{\mathbf{k}})-\sum^{\,}_{(\mathbf{k}^{\,}_1\mathbf{k}^{\,}_2\mathbf{q}^{\,}_1)}H^{\,}_{\mathbf{k}^{\,}_1\mathbf{k}^{\,}_2\mathbf{q}^{\,}_1}\delta^{\,}_{\{\mathbf{k}^{\,}_1,\mathbf{k}^{\,}_2\}\mathbf{k'}}\nonumber\\
&-\sum^{\,}_{\substack{(\mathbf{k}^{\,}_1\mathbf{k}^{\,}_2\mathbf{q}^{\,}_1)\\ \mathbf{k}^{\,}_j\neq\mathbf{k'}(j=1,2)}}H^{\,}_{\mathbf{k}^{\,}_1\mathbf{k}^{\,}_2\mathbf{q}^{\,}_1}\delta^{\,}_{\{-\mathbf{k}^{\,}_1+\mathbf{q}^{\,}_1,-\mathbf{k}^{\,}_2-\mathbf{q}^{\,}_1\}-\mathbf{k}+\mathbf{q}}\nonumber\\
&-\hspace*{-1cm}\sum^{\,}_{\substack{(\mathbf{k}^{\,}_1\mathbf{k}^{\,}_2\mathbf{q}^{\,}_1)\\ \mathbf{k}^{\,}_j\neq\mathbf{k'}(j=1,2)\\ \qquad-\mathbf{k}^{\,}_1+\mathbf{q}^{\,}_1,-\mathbf{k}^{\,}_2-\mathbf{q}^{\,}_1\neq-\mathbf{k}+\mathbf{q}}}\hspace*{-1.2cm}H^{\,}_{\mathbf{k}^{\,}_1\mathbf{k}^{\,}_2\mathbf{q}^{\,}_1}\delta^{\,}_{\{-\mathbf{k}^{\,}_1+\mathbf{q}^{\,}_1,-\mathbf{k}^{\,}_2-\mathbf{q}^{\,}_1\}-\mathbf{k'}-\mathbf{q}},
\end{align}
which is $E^{\,}_3-E^{\,}_0$ as expected from Eqs.\,(\ref{energy0}) and (\ref{energy3}).

{\it Superfluid fraction.---}We now apply the theory to a practical calculation, which involves the computation of the superfluid fraction $f^{\,}_s=n^{\,}_s/n$, with $n^{\,}_s$ the superfluid density and $n$ the total density. To achieve that, we calculate the current response $\braket{\hat{\mathbf{j}}(\mathbf{q})}$ to a transverse vector potential $\mathbf{A}(\mathbf{q})$. The current density $\hat{\mathbf{j}}(\mathbf{q})$ has two components, the paramagnetic component %$\hat{\mathbf{j}}^{\,}_1(\mathbf{q})=\frac{e\hbar}{m}\sum^{\,}_\mathbf{k}(\mathbf{k}-\mathbf{q}/2)(c^\dagger_{\mathbf{k}-\mathbf{q}\uparrow}c^{\,}_{\mathbf{k}\uparrow}-c^\dagger_{\mathbf{-k}\downarrow}c^{\,}_{\mathbf{-k}+\mathbf{q}\downarrow})$ 
\begin{align}
    \hat{\mathbf{j}}^{\,}_1(\mathbf{q})=\frac{e\hbar}{m}\sum^{\,}_\mathbf{k}(\mathbf{k}-\frac{\mathbf{q}}{2})(c^\dagger_{\mathbf{k}-\mathbf{q}\uparrow}c^{\,}_{\mathbf{k}\uparrow}-c^\dagger_{\mathbf{-k}\downarrow}c^{\,}_{\mathbf{-k}+\mathbf{q}\downarrow}),
\end{align}
and the diamagnetic component  %$\hat{\mathbf{j}}^{\,}_2(\mathbf{q})=-\frac{e^2}{m\mathcal{V}}\sum^{\,}_\mathbf{k}\mathbf{A}(\mathbf{q}-\mathbf{k})\hat{\rho}(\mathbf{k})$, 
\begin{align}
\hat{\mathbf{j}}^{\,}_2(\mathbf{q})=-\frac{e^2}{m\mathcal{V}}\sum^{\,}_\mathbf{k}\mathbf{A}(\mathbf{q}-\mathbf{k})\hat{\rho}(\mathbf{k}),
\end{align}
with $\mathcal{V}$ the volume of the system and the density operator $\hat{\rho}(\mathbf{k})=\sum^{\,}_{\mathbf{k'}\sigma'}c^\dagger_{\mathbf{k'}-\mathbf{k}\sigma'}c^{\,}_{\mathbf{k'}\sigma'}$. Because $\braket{\hat{\rho}(\mathbf{k})}=N\delta^{\,}_{k0}$ with $N$ the total particle number, we have the diamagnetic current $\braket{\hat{\mathbf{j}}^{\,}_2(\mathbf{q})}=-\frac{ne^2}{m}\mathbf{A}(\mathbf{q})$. As to the paramagnetic current, it can be calculated by the linear response theory or, equivalently, the second-order perturbation theory  
\begin{align}\label{j1}
\braket{\hat{\mathbf{j}}^{\,}_1(\mathbf{q})}=\sum^{\,}_{\ell (\neq 0)}\frac{[\braket{\ell|\hat{\mathbf{j}}^{\,}_1(\mathbf{-q})|\Psi}]^*\braket{\ell|\hat{H}^{\,}_1|\Psi}}{E^{\,}_0-E^{\,}_\ell} + (\mathbf{q}\rightarrow\mathbf{-q})^*,
\end{align}
where $\hat{H}^{\,}_1=-\frac{1}{\mathcal{V}}\sum^{\,}_\mathbf{q'}\hat{\mathbf{j}}^{\,}_1(\mathbf{-q'})\cdot\mathbf{A}(\mathbf{q'})$, $\ket{\ell}$ is a symbolic representation of a four-particle excited state $c^\dagger_{\mathbf{k}+\mathbf{q}\uparrow}\ket{\mathbf{k}\mathbf{k'}\mathbf{q'}}$, $E^{\,}_0$ is the ground state energy, and $E^{\,}_\ell$ is the excited state energy associated with $\ket{\ell}$. Then within the DQA, after some analysis we have %$\bra{\ell}(c^\dagger_{\mathbf{k}^{\,}_1+\mathbf{q}^{\,}_1\uparrow}c^{\,}_{\mathbf{k}^{\,}_1\uparrow}-c^\dagger_{-\mathbf{k}^{\,}_1\downarrow}c^{\,}_{-\mathbf{k}^{\,}_1-\mathbf{q}^{\,}_1\downarrow})\ket{\Psi} = \delta^{\,}_{\mathbf{q}^{\,}_1\mathbf{q}}(\delta^{\,}_{\mathbf{k}^{\,}_1\mathbf{k}}\alpha^{\,}_{\mathbf{k}\mathbf{k'}\mathbf{q'}} + \delta^{\,}_{\mathbf{k}^{\,}_1,\mathbf{k'}-\mathbf{q}}\alpha^{\,}_{\mathbf{k}+\mathbf{q},\mathbf{k'}-\mathbf{q},\mathbf{q'}+\mathbf{q}} - \delta^{\,}_{\mathbf{k}^{\,}_1,\mathbf{k}-\mathbf{q'}}\alpha^{\,}_{\mathbf{k}+\mathbf{q},\mathbf{k'},\mathbf{q'}} - \delta^{\,}_{\mathbf{k}^{\,}_1,\mathbf{k'}+\mathbf{q'}}\alpha^{\,}_{\mathbf{k}+\mathbf{q},\mathbf{k'},\mathbf{q'}+\mathbf{q}})$ 
\begin{align}
\bra{\ell}(c^\dagger_{\mathbf{k}^{\,}_1+\mathbf{q}^{\,}_1\uparrow}&c^{\,}_{\mathbf{k}^{\,}_1\uparrow}-c^\dagger_{-\mathbf{k}^{\,}_1\downarrow}c^{\,}_{-\mathbf{k}^{\,}_1-\mathbf{q}^{\,}_1\downarrow})\ket{\Psi}\nonumber\\
= \delta^{\,}_{\mathbf{q}^{\,}_1\mathbf{q}}\big(\delta^{\,}_{\mathbf{k}^{\,}_1\mathbf{k}}&\alpha^{\,}_{\mathbf{k}\mathbf{k'}\mathbf{q'}} + \delta^{\,}_{\mathbf{k}^{\,}_1,\mathbf{k'}-\mathbf{q}}\alpha^{\,}_{\mathbf{k}+\mathbf{q},\mathbf{k'}-\mathbf{q},\mathbf{q'}+\mathbf{q}}\nonumber\\
- \delta^{\,}_{\mathbf{k}^{\,}_1,\mathbf{k}-\mathbf{q'}}&\alpha^{\,}_{\mathbf{k}+\mathbf{q},\mathbf{k'},\mathbf{q'}} - \delta^{\,}_{\mathbf{k}^{\,}_1,\mathbf{k'}+\mathbf{q'}}\alpha^{\,}_{\mathbf{k}+\mathbf{q},\mathbf{k'},\mathbf{q'}+\mathbf{q}}\big),
\end{align}
and therefore the proportional relation
\begin{align}\label{proportionality}
\braket{\ell|\hat{\mathbf{j}}^{\,}_1(\mathbf{-q})|\Psi}\propto \bigg[(\mathbf{k}+\frac{\mathbf{q}}{2})\alpha^{\,}_{\mathbf{k}\mathbf{k'}\mathbf{q'}} + (\mathbf{k'}-\frac{\mathbf{q}}{2})\alpha^{\,}_{\mathbf{k}+\mathbf{q},\mathbf{k'}-\mathbf{q},\mathbf{q'}+\mathbf{q}}\nonumber\\ 
-(\mathbf{k}-\mathbf{q'}+\frac{\mathbf{q}}{2})\alpha^{\,}_{\mathbf{k}+\mathbf{q},\mathbf{k'},\mathbf{q'}} - (\mathbf{k'}+\mathbf{q'}+\frac{\mathbf{q}}{2})\alpha^{\,}_{\mathbf{k}+\mathbf{q},\mathbf{k'},\mathbf{q'}+\mathbf{q}}\bigg],
\end{align}
which vanishes in the long wavelength limit $\mathbf{q}\rightarrow 0$. Because we have $E^{\,}_\ell-E^{\,}_0>0$ in Eq.\,(\ref{j1}), immediately we see that $\braket{\hat{\mathbf{j}}^{\,}_1(\mathbf{q})}\rightarrow 0$ when $\mathbf{q}\rightarrow 0$ and therefore the London equation $\braket{\hat{\mathbf{j}}(\mathbf{q})}=-\frac{ne^2}{m}\mathbf{A}(\mathbf{q})$, together with the superfluid fraction $f^{\,}_s=1$ same as in conventional BCS superconductors. Note that our result differs from the suppressed superfluid fraction previously obtained using solvable models \cite{Wang2022,Hu2024}, either because one model has a gapless Fermi surface \cite{Wang2022} or the other (effectively) breaks the translational invariance \cite{Hu2024}.

{\it Conclusion and discussion.---}The BCS theory is a milestone in the development of superconductivity and superfluidity. In this paper, following steps of BCS, we constructed an effective theory based on the condensation of fermion quartets, instead of Cooper pairs, and applied it to the calculation of the superfluid fraction. Because of the interacting nature of the fermion quartet problem, we adopted an approximation that we call the ``dilute quartet approximation" during the construction, with a focus on the dilute quartet regime. We found that the superfluid fraction is the same as in conventional superconductors, although previous works reached a different conclusion \cite{Wang2022,Hu2024}.

Now we discuss a little bit the case that goes beyond the dilute quartet regime by setting the chemical potential to the Fermi energy. Due to the constraint of the Fermi surface, we have the variational wavefunction 
\begin{align}\label{w.f.2}
\ket{\Psi'}=\frac{1}{\sqrt{\mathcal{N}^{'}}} \prod^{\,}_{\{\mathbf{k}^{\,}_1,\mathbf{k}^{\,}_2\}}(1+\alpha^{\,}_{\mathbf{k}^{\,}_1\mathbf{k}^{\,}_2}c^\dagger_{\mathbf{k}^{\,}_1\uparrow}c^\dagger_{-\mathbf{k}^{\,}_1\downarrow}c^\dagger_{\mathbf{k}^{\,}_2\uparrow}c^\dagger_{-\mathbf{k}^{\,}_2\downarrow})\ket{0},
\end{align}
which is a specialization of Eq.\,(\ref{w.f.}) with $\mathbf{q}=0$ [see also Fig.\,\ref{Fig_Binding} (b)]. If we enforce previous procedures and develop an effective theory in parallel, we would get an analogous expression for the variational parameter $\alpha^{\,}_{\mathbf{k}^{\,}_1\mathbf{k}^{\,}_2}=(\xi^{\,}_{\mathbf{k}^{\,}_1\mathbf{k}^{\,}_2}-E^{\,}_{\mathbf{k}^{\,}_1\mathbf{k}^{\,}_2})/\Delta^*_{\mathbf{k}^{\,}_1\mathbf{k}^{\,}_2}$, with $\xi^{\,}_{\mathbf{k}^{\,}_1\mathbf{k}^{\,}_2}$, $E^{\,}_{\mathbf{k}^{\,}_1\mathbf{k}^{\,}_2}$, and $\Delta^{\,}_{\mathbf{k}^{\,}_1\mathbf{k}^{\,}_2}$ similarly defined as before. Then we have $|\alpha^{\,}_{\mathbf{k}^{\,}_1\mathbf{k}^{\,}_2}|\simeq 1$ around the Fermi surface, which violates the dilute quartet limit ($|\alpha^{\,}_{\mathbf{k}^{\,}_1\mathbf{k}^{\,}_2}|\ll 1$). Thus, how to develop a new approximation scheme to tackle the problem is an open question. Furthermore, whether there exists a crossover from Eq.\,(\ref{w.f.}) to Eq.\,(\ref{w.f.2}) is another open question, since other electronic orders such as the conventional superconducting phase may intervene \cite{Yao2017}. We leave both questions for future investigation.

%\acknowledgments
{\it Acknowledgments.---}C.W. is supported by the National Natural Science Foundation of China under the Grant No. 12234016 and the New Cornerstone Science Foundation.

\end{document}